# Exploitees vs. Exploiters: Dynamics of Exploitation

*Davood Qorbani* (*Twitter & LinkedIn @DavoodQorbani*)
Norwegian University of Science and Technology, Norway

**Abstract**

Either saying that the market is structured to promote workforce exploitation in some sections or the people who participate there benefit from the existing structure and exploit, exploitation happens – systematically or opportunistically. This research presents a perspective on how workforce exploitation may occur when vulnerable groups voluntarily seek employment in the labor market.

## 1. Introduction

**A close body of research from ecology and natural sciences**

The interaction between two species could take different forms in nature and the environment. Among different types of interaction that exist there, collaborations, cohabitation/coexistence, mutually exploitations, and prey vs. predator are the interests of this paper, as they create a foundation for further elaboration and discussion of exploitation.

Examples of *collaboration* could be a cleaner fish that cleans the mouth of a larger fish from bacteria (Dixon & Smith, 2009) or a bird that cleans the teeth of a crocodile. In both examples, the former animal is being fed, and the latter one avoids cavities as a consequence of this collaboration. An example of *coexistence* is observed between Meat Ants and a type of caterpillars. The latter discharges a sugary fluid that those ants consume, and in exchange, the ants protect caterpillars from their predators. "Ravens Guide Wolves to prey" can be categorized as both collaboration and mutual exploitations of an opportunity, an excellent cooperation type, especially during harsh winters. Wolves hunt down prey, eat and get fed, then leave the scene, and ravens enjoy the feast of leftovers (Sfarra, 2015). The last category of interest is prey vs. predator. This is a more common type of interaction in nature. Wolves vs. deer or sheep, sharks vs. smaller fish, and even cannibalism among some species in which the stronger preys on the weaker (see Berryman, 1992; Gross & Sturis, 1992; Preisser et al., 2005; Swart, 1990; Toro & Aracil, 1988). In all these examples, more or less, the behavior is derived from instincts and the need to survive.

In such situations, the prey or the weaker party is not aware of the experienced exploitation by its common sense.

**Exploitees versus Exploiters**

Compared to animals in nature, human beings feel and sense treatments (phenomena) such as discrimination or exploitation. Even monkeys dislike unfair pay – see de Waal (2013). There are abundant documents on how different human societies and communities have suffered and bore the scars of exploitation in various corners of Earth, such as North America and Africa. Whether in the past or modern times, "individuals from particular groups are inherently exploitable as a result of systemic unjust social conditions" (Deranty, 2016, p. 33).

Real-world exploitation happens in various forms, such as underpayment or questionable and poor working conditions (see Pattisson et al., 2021; Rypeng & Iversen, 2022). Official figures estimated that 40.3 million people are in modern slavery, among them 24.9 million in forced labor: 16 million people are exploited in the private sector such as domestic work, construction, or agriculture; 4.8 million in forced sexual exploitation, and 4 million in forced labor imposed by state authorities (ILO, 2017). One often-overlooked type of exploitation also happens in some academic environment where conditions for receiving funding bitters the life of researchers, and sacrifices becomes necessary to climb up the career ladder; then, health problems such as anxiety attacks, migraines, and insomnia (Hugo, 2017); psychological distress; psychiatric disorder; and mental health problems (Levecque et al., 2017) appears.

Wertheimer (1999) took US student-athletes and the sport-related services they potentially offer and talents they (may) possess as the basis of one of his cases and developed an entire book chapter around it. To serve the purpose of this paper, a synthesized and adjusted summary of his exemplary book, *Exploitation,* is as follows*:*

One claim could be that those students generate "nonfinancial benefits for which they are not compensated" (p. 84). They provide services in the form of research and projects they do, which finally adds to the reputation of the university or the research center they are affiliated. From one perspective, it is reasonable to argue that other students, faculty, and alumni benefit from that generated reputation (obviously, reputation is among the factors that attract new students). The public also benefits from the incremental flow of knowledge which is being generated.

> *In these cases, it is not that the universities are hiding financial revenues. Rather, the [students…] create important benefits that are not converted into financial*

> *revenues. [...] In any case, this suggests that moral criteria are required not only to evaluate the distribution of a social surplus but to establish the presence and magnitude of a social surplus. ...*
>
> *Let us also assume that B contributes to the competitive and (therefore) financial success of the [...] program. Is B exploited? And, if so, is it a case of harmful exploitation or mutually advantageous [and consensual] exploitation? [Before that, the reader] should understand such benefits in terms of their ex-ante value rather than their ex-post values. [...] In one sense, educational opportunities are like lottery tickets. There is no guarantee that anyone admitted to an institution will succeed [...]. Whether B actually receives an education or a degree depends largely, although not altogether, upon whether B makes the appropriate effort. The ex-post value of the educational opportunity is at least somewhat subject to B's control. Thus, the question is not whether B actually receives an education or graduates but specifying the ex-ante value of the educational resources or opportunities that are provided to B. [...] The value of educational opportunities to B is, of course, not only a function of B's efforts but also B's capacity to use those educational opportunities.*
>
> *It might be replied that A does not exploit B if B voluntarily assumes the risk of not benefiting from the educational opportunities provided by A. But this is wrong. <u>A can exploit B even if B consents to the exploitation</u>. [...]. Second, it is not clear that this exploitation is genuinely consensual* (Wertheimer, 1999, pp. 84-86).

Wertheimer closed that chapter by concluding that a systemic form of exploitation does exist in the context mentioned. However, he primarily blamed society at large as the responsible party for the background conditions, rather than taking universities to be "solely or even primarily to blame" (Wertheimer, 1999, p. 96). In a paper, Deranty (2016) mentioned Wertheimer's arguments and distinguished between opportunistic versus systemic exploitation. He argued that while opportunistic exploitation is inevitable and exists everywhere contingently (or case by case), much emphasis shall be put on systemic exploitations because they affect an entire population or a group.

In the case of exploitation, the exploitee has something to offer, such as work capacity or intangible skills which potentially generate revenue. Then, there will be a division of revenues between the employee and the employer. Depending on the level and the stage of employment, there could be traces or different levels of exploitation. However, the chances of exploitation are much higher in the lower employment levels, bearing in mind the bargaining power of either side.

It is also worth mentioning the presence of a unique dynamic in the relation of exploitee-exploiter. Throughout the history of humankind, there has been a vicious cycle in place: a reinforcing loop that says the richer someone is, the richer he can get, and vice versa. The poorer someone is, the poorer he is going to be or may become. In other words, people become locked into a mode of behavior in which the rich get richer, and the poor remain poorer. The more someone has, the more he gets or earns, and the more he gets, the more he has the next time – similar to success to the successful archetype, see Senge (2006).

**Counterfactual arguments: Is exploitation opportunistic or systemic?**

According to Deranty (2016), systemic exploitation exists if people are exploited because of their vulnerability due to being a member of a specific community or an ethnic group or sharing common traits. As mentioned earlier, he gave priority to addressing systemic over opportunistic exploitation. However, he favored "social accounts of exploitation, […] which mix opportunistic and systemic exploitation. Opportunistic exploitation might well share conceptual content with systemic exploitation, or it might derive its content from systemic exploitation by analogy or by metaphor… "(Deranty, 2016, p. 33). The unanswered ambiguity is to find the common denominator of the three types of exploitations, i.e., opportunistic vs. systemic vs. social, as depicted in Figure 1.

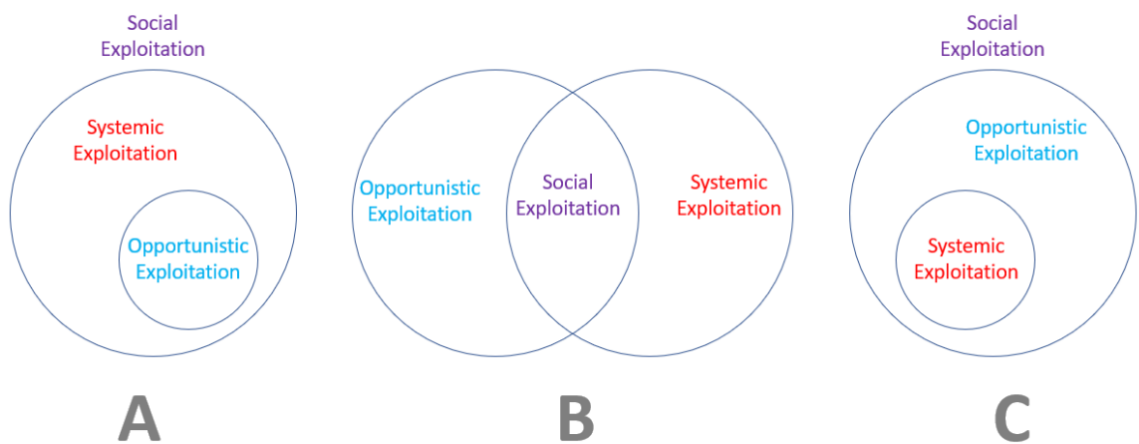

*Figure 1: Opportunistic vs. Systemic vs. Social Exploitation*

It is a challenge to answer this question; however, one may argue that exploitation seems to be a social type that dominates both exploitees and exploiters. Furthermore, the social account of exploitation is not a mix but a synergy between opportunistic and systemic exploitation: powerful reinforcing and balancing feedback loops synergize and lockdown both exploitees and exploiters.

## 2. Elaboration of Exploitation using a Causal Loop Diagram

This section benefits from the causal loop diagramming (cf. Meadows & Wright, 2009; Morecroft, 2015; Senge, 2006; Sterman, 2000) to conceptually outline a generic platform within which exploitation may occur when vulnerable groups voluntarily seek employment in the labor market – in the absence or failure of regulatory measures that support the weaker side. Boundaries and a level of abstract are considered to provide insights on the topic while keeping the diagram accessible, understandable, and readable. While there could be various narrations of exploitation in the labor market, Figure 2 is an attempt to be a generic explanation.

In many circumstances, the exploitation has not reached its limit, i.e., the limit on the supply of those available to be exploited. Very often, there is still a large pool of people willing to be exploited, and perhaps that is the question of price: what are the ex-ante and the ex-post value of unfilled positions?

Since there are many uncertainties about the future, anticipating the ex-ante value of something is likely overestimated. That uncertainty, combined with some signals from the market (in its general meaning), could whet the appetite of potential exploitees. One signal could be the offered salary for positions. But how?

Let us begin with loop B1 (highlighted in red) in Figure 2. When the pool of potential (susceptible) exploitees who are available/willing to be exploited is reduced, then the price of exploitation increases, because potential exploitees demand higher salaries. Consequently, the exploiter needs to offer a higher wage to maintain or increase the relative attractiveness of his offer compared to the alternative options in the market. After that, there are two other effects:

First, an increased salary could increase the relative attractiveness of the target work environment compared to alternative jobs, which subsequently could increase the ex-ante value of such positions. The word "relative attractiveness" is a keyword that is further discussed in this research. Such increased ex-ante value leads to a higher number of vacancies to be filled.

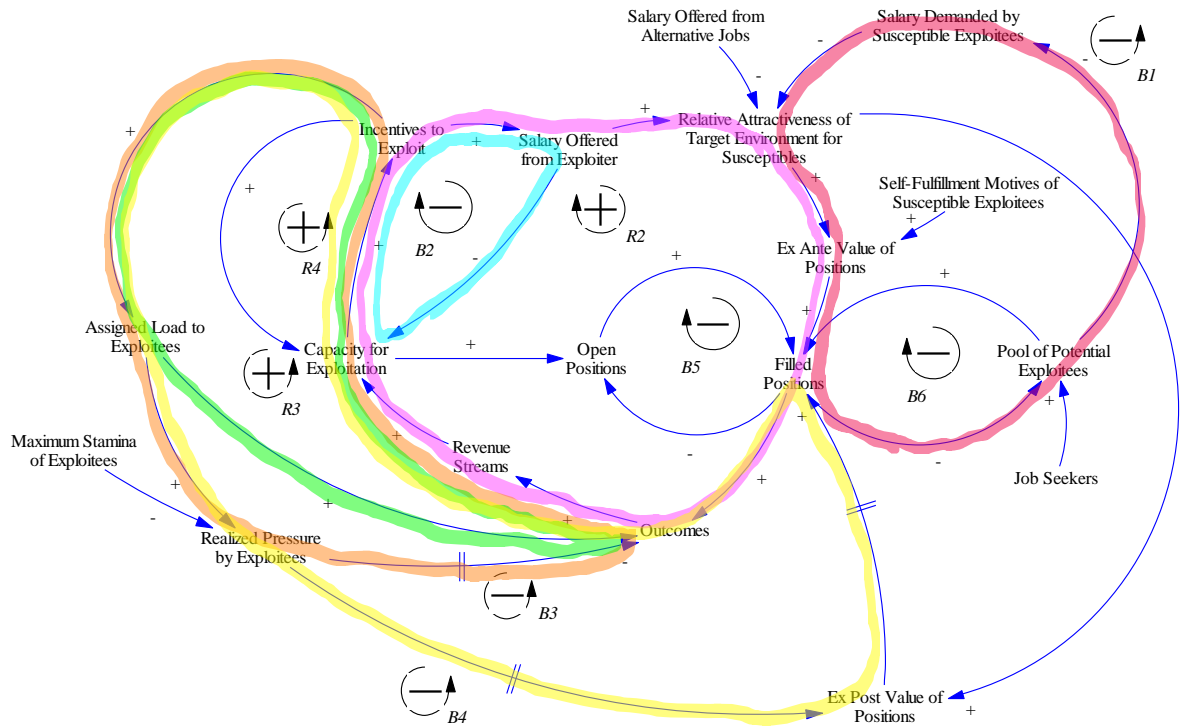

*Figure 2: A Causal Loop Diagram of Exploitation in the Labor Market*
Abbreviations in figure → B: a balancing loop; R a reinforcing loop; sign ‖ : a delay between cause and its effect

Consequently, more outcomes will be produced, generating higher revenue streams, which in turn contributes to the capacity for exploitation. Higher exploitation capacity leads to higher incentives to exploit and triggers offerings of higher salaries. This is a reinforcing loop, R2 (the loop in pink). However, the increased salary drains or limits the capacity for exploitation that the exploiter could build upon (B2, the loop in blue). Because the exploiters may not have sufficient funds to keep on and conduct the exploitation, this forms a balancing loop that counteracts the former loop.

There are other essential dynamics here. When incentives to exploit increase, it is highly probable that the assigned load to exploitees increases. There will be short- and long-term effects here. In the short term (R3, the loop in green, a reinforcing one), the outcomes increase, which leads to higher revenues for the exploiter. However, the realized pressure by exploitees in the long term (B3, the loop in orange, a balancing one) will probably show itself in the form of exhaustion and burn-out (not included in the diagram). Consequently, there will be declines in the outcomes. The other notable consequence is that the ex-post value of such positions will suffer (B4, the loop is yellow), probably by the negative impact of word of mouth and complaints from current exploitees. Such dynamics hinder the process of filling open vacancies. These are some of the significant loops that shape the dynamics of exploitation.

## 3. Further discussion

The issue elaborated here also connotates a labor-market dynamic problem in which the price of the commodity of interest (i.e., the average price of people who are exploited) is determined in a negotiation between those who offer the salary (the exploiters) and those who earn the salary (the exploitees). We shall distinguish between the salary offered, and the salary demanded. Such differentiation is essential because both are inputs to estimate the relative attractiveness mentioned earlier. When the pool of potential exploitees shrinks, then the demanded salary will be higher and vice versa. If that pool increases, the salary demanded gradually becomes equal to the salary offered. In that case, only a fraction (e.g., one-half) of those available for exploitation would let themselves be exploited at the salary offered.

Various scenarios are plausible depending on the larger economy's macroeconomic variables. For simplicity and understandability of the narration, let us assume that if the offered and the demanded salaries are equal, only half of the potential exploitees will be willing to let themselves be exploited at that average price. Because, at its best, that average salary is only satisfactory to half of the mentioned population. This situation is mapped on the distribution curve of Figure 3, where the X-axis is the salary offered. In the middle is the median, and if we go beyond that point, we can include more people willing to let themselves be exploited at a higher price.

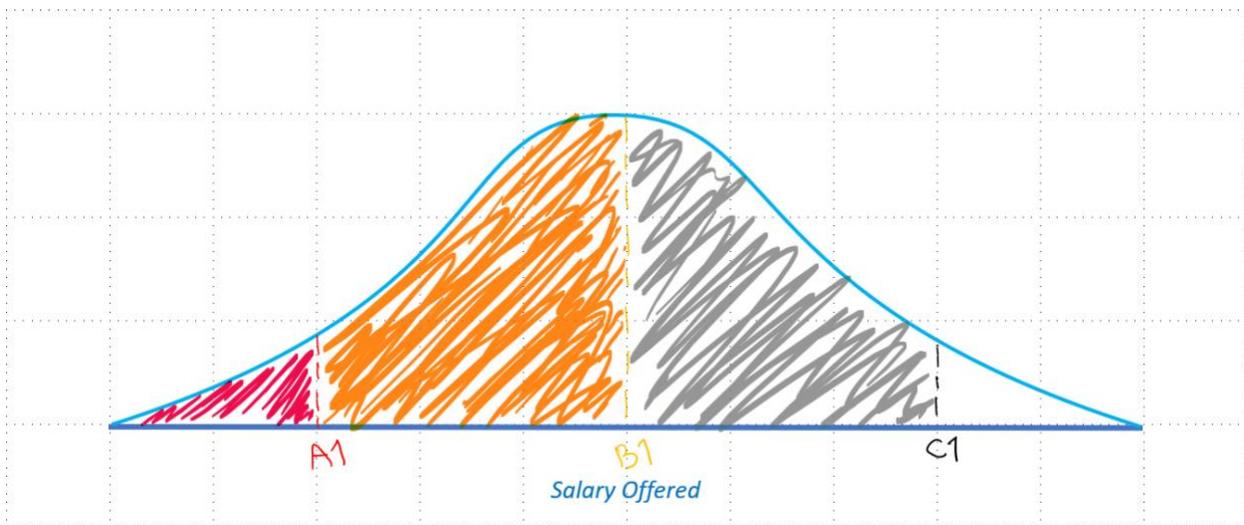

*Figure 3: A Normal Distribution Graph for Relative Attractiveness of Target Environment*

Suppose that the salary offered is relatively set low (point A1); then, only a tiny fraction of those potentials who are under that curve will find it a satisfactory salary and let themselves be exploited. If the salary is increased to point B1, half of the mentioned population will be willing to be exploited (and the rest will not). At point C1, assume that 90 percent will be willing to be exploited (equals to the area under the curve up until point C1), and so on.

The accumulative graph in Figure 4 also can be considered an alternative mapping tool, since the interpretation may sound easier to explain to a broader audience. Like the previous graph, here we compare the salary offered with the average salary asked for. Thus, the ratio between two becomes a number between zero and one (100%). Almost everyone lets themselves be exploited at the latter value because the exploiter has offered a very high salary, and the value becomes near zero if a meager salary is offered. For instance, point B2 in the accumulative graph of Figure 4 corresponds to point B1 in the normal distribution graph of Figure 3, at which fifty percent of the potential (susceptible) exploitees would be willing to be exploited.

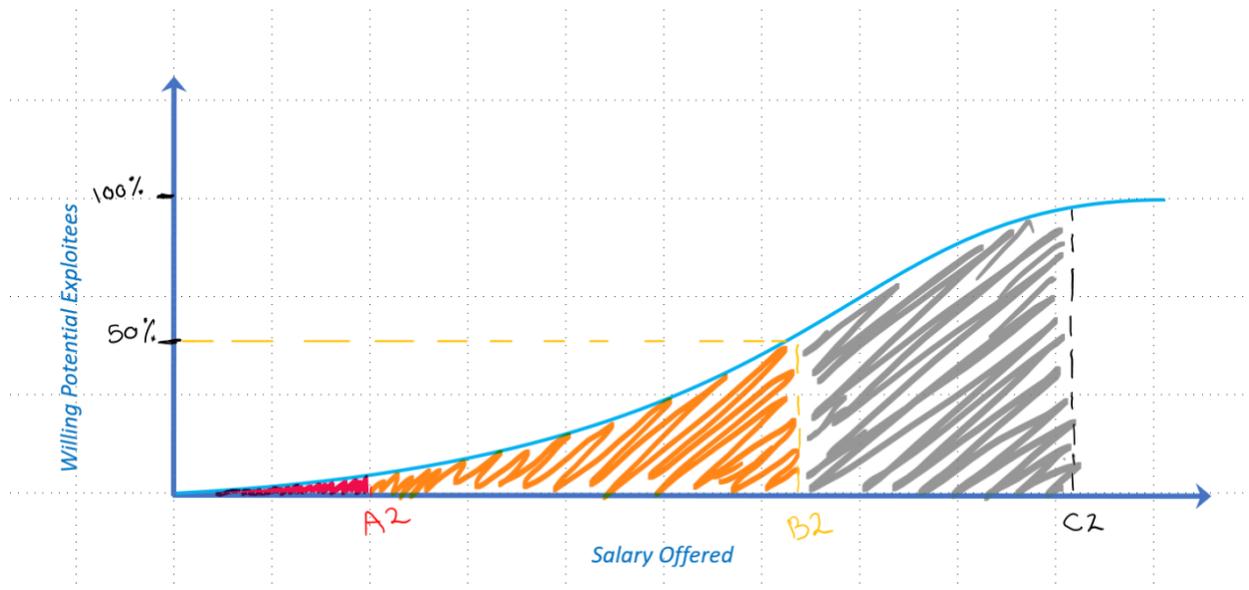

*Figure 4: An Accumulative Graph for Relative Attractiveness of Target Environment*

Now let us have a flashback to a reconstructed normal distribution in Figure 5: A parallel and comparable example that better clarifies our argument is prostitutes. Street prostitutes, who are at the lower end in terms of the price they demand – and do that activity for a living – are probably accessible to exploiters at lower prices (e.g., point A1). At the other end are high-end males or females who do the activity to earn money on the side. The latter category demand higher prices, and the exploiter needs to offer significant prices to include them in their susceptible exploitees cohort (e.g., point C1). This premise can be assumed in many other contexts.

Obviously, there exist biases. For instance, there could be an intense increase at the beginning of the distribution, i.e., a skewed one, as in Figure 5. It means that in the corresponding accumulative graph, Figure 6, there will be a rapid rise in the fraction of exploitees willing to be exploited, which takes a very high price to include the final uninterested potential exploitees, willing to let themselves be exploited.

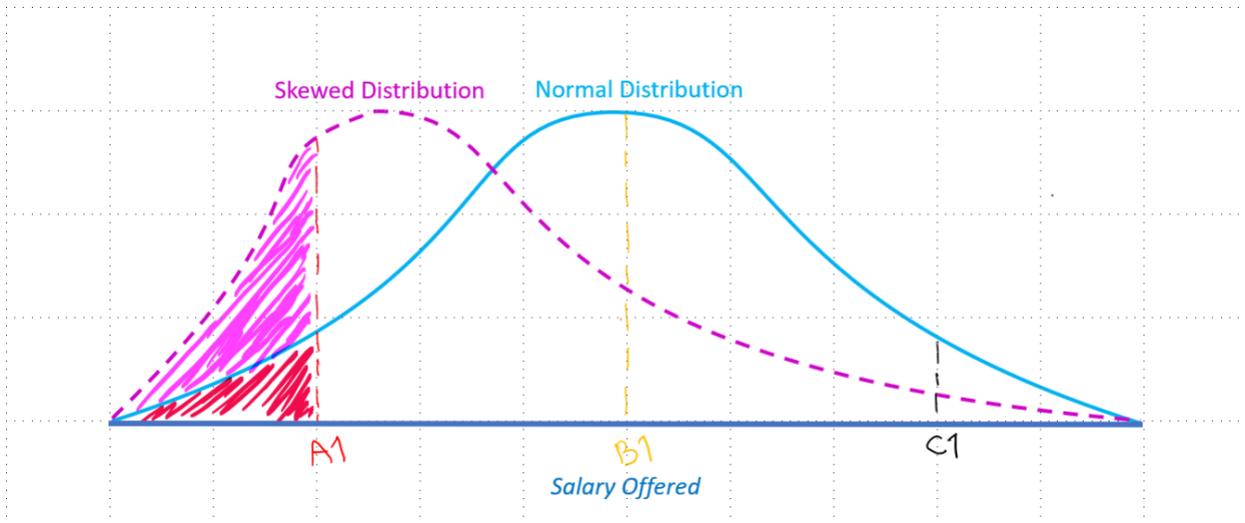

*Figure 5: A Skewed Distribution for Relative Attractiveness of Target Environment*

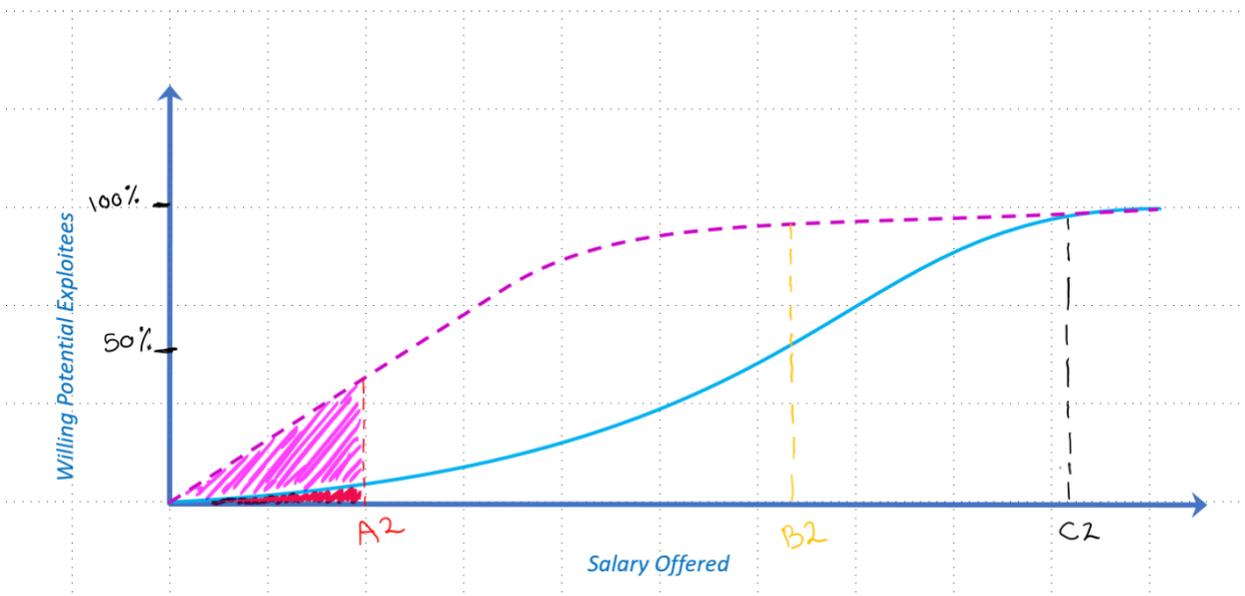

*Figure 6: Normal and Biased Accumulative Graph for Relative Attractiveness of Target Environment*

That curve possibly has different shapes, which yield different fractions for every input, consequently affecting the ex-ante value. Multiplying that value – the effect of the salary and other self-fulfillment factors on the willingness of potential exploitees to let themselves be exploited – by the pool of susceptible exploitees, gives the current number of exploitees (or filled positions).

Exploitees vs. Exploiters: *Dynamics of exploitation*

The above arguments are an essential foundation for the discussion in this paper because the distribution of *desperation* among people has ample importance. If many desperate people struggle to earn a living, those people would probably be willing to do anything. On the other hand, most people in rich countries have a relatively sufficient income nowadays, and relatively few people will be on the verge of letting themselves be exploited. They might have a luxury need or a fetish that is exposing them to exploitation. However, those people might not be a large part of the population willing to be exploited.

"We shape our buildings; thereafter, our buildings shape us – Winston Churchill" (Sterman, 2000, p. 137). A social system can be corrected from the inside if appropriate decision rules are implemented wisely and timely. Until then, the greed of exploiters and the various vulnerabilities of exploitees feed the dynamics of exploitation.

## 4. Future direction

This paper can shape the first part of a multi-level investigation on the topic of exploitation in the labor market from a systems perspective. A quantitative System Dynamics model that includes price elasticities and income distribution of target groups helps to develop viable policy options while limiting unintended negative consequences. For instance, imposing minimum wage laws cause the unintended consequence of increasing unemployment as employers react to the price increase and lower their demand. Such policies can be tested in a quantified model.